\documentclass[%
 reprint,
superscriptaddress,
 amsmath,amssymb,
 pra
]{revtex4-1}

\usepackage{graphicx}
\usepackage{subfigure} 
\usepackage{epstopdf} 
\usepackage{hyperref}

\usepackage{dcolumn}
\usepackage{bm}

\begin{document}

\preprint{APS/123-QED}

\title{Lifetimes of ultra-long-range strontium Rydberg molecules}

\author{F. Camargo}
	\email{Francisco.Camargo@Rice.edu}
\author{J.\,D. Whalen}%
\author{R. Ding}%
 
\affiliation{
 Department of Physics \& Astronomy, Rice University, Houston, TX 77251, USA
}


\author{H.\,R. Sadeghpour }
\affiliation{
ITAMP, Harvard-Smithsonian Center for Astrophysics, Cambridge, MA 02138, USA
}%

\author{S. Yoshida}
\author{J. Burgd\"orfer}%
\affiliation{%
Institute for Theoretical Physics, Vienna University of Technology, Vienna, Austria, EU
}%

\author{F.\,B. Dunning}%
\author{T.\,C. Killian}%
\affiliation{
 Department of Physics \& Astronomy, Rice University, Houston, TX 77251, USA
}

\date{\today}

\begin{abstract}
The lifetimes of the lower-lying vibrational states of ultralong-range strontium Rydberg molecules comprising one ground-state $5\textup{s}^2~ ^1 \textup{S}_0$ atom and one Rydberg atom in the $5\textup{s}38\textup{s}~ ^3\textup{S}_1$ state are reported.  The molecules are created in an ultracold gas held in an optical dipole trap and their numbers determined using field ionization, the product electrons being detected by a microchannel plate.  The measurements show that, in marked contrast to earlier measurements involving rubidium Rydberg molecules, the lifetimes of the low-lying molecular vibrational states are very similar to those of the parent Rydberg atoms.  This results because the strong p-wave resonance in low-energy electron-rubidium scattering, which plays an important role in determining the molecular lifetimes, is not present for strontium.  The absence of this resonance offers advantages for experiments involving strontium Rydberg atoms as impurities in quantum gases and for testing theories of molecular formation and decay. 
\end{abstract}

\pacs{Valid PACS appear here}
\maketitle


\section{\label{sec:level1}Introduction}
Scattering of the excited electron in a Rydberg atom from a neighboring ground state atom can bind the two together to form an ultralong-range Rydberg molecule.  The interaction between the excited electron and ground state atom can be described using a Fermi pseudopotential that includes both s- and p-wave scattering contributions resulting in a molecular potential that can bind multiple vibrational states with binding energies typically of a few to a few tens of MHz.  Such molecules are of interest because they represent a new mechanism for chemical binding and because they possess unusual physical characteristics.  For example, they are very large with internuclear separations comparable to the size of the parent Rydberg atoms, i.e., ${\sim}2n^2$ a.u., and they can have large permanent electric dipole moments, ${\sim} n^2$ a.u., even in the case of a homonuclear molecule.

	While the existence of Rydberg molecules was originally predicted theoretically \cite{PhysRevLett.85.2458}, they have now been observed using a variety of species including rubidium, cesium, and strontium \cite{bendkowsky2009observation, li2011homonuclear, PhysRevLett.109.173202, PhysRevA.92.031403}.  Initial experimental studies involved spherically-symmetric Rb(ns) states and the creation of molecules with small electric dipole moments.  Measurements, however, have now been extended to include anisotropic P and D Rydberg states \cite{PhysRevLett.111.053001, PhysRevLett.114.133201, PhysRevLett.112.143008, PhysRevLett.112.163201} and, using Cs(ns) Rydberg states, to the creation of so-called “trilobite” states with very large permanent electric dipole moments \cite{booth2015production}.  In addition, molecules comprising one Rydberg atom and up to four ground state atoms have been observed \cite{gaj2014molecular}.
	
	One interesting question concerning Rydberg molecules relates to their lifetimes and how these compare to the lifetimes of the parent Rydberg atoms especially in a dense background gas.  Earlier studies \cite{0953-4075-44-18-184004} using rubidium Rydberg molecules demonstrated that the natural lifetime of the parent Rb(35s) atoms, $63(8)~\mu s$, is significantly longer than that of the molecules and that the lifetime of the molecules themselves depends on their vibrational states, the lifetime of the $\nu=1$ state, $30(11)~\mu s$, being substantially shorter than that of the ground $\nu=0$ state, $48(9)~\mu s$.  This behavior was attributed to the presence of a strong p-wave shape resonance in low-energy electron scattering from rubidium, which causes a sharp decrease (step) in the molecular potential for internuclear separations, $R$, (for 35s states) below ${\sim} 1200~\textup{a.u.}$  Rubidium molecular states bound by quantum reflection at this step or in local minima at larger internuclear separations can still penetrate to short range through imperfect reflection or tunneling, leading to a reduced natural lifetime. A similar p-wave resonance exists for cesium, although its effects on molecular lifetimes have not been examined. A significant density dependence in the loss rates for rubidium molecules was also observed, the cross section for loss in collisions with neutral rubidium atoms being comparable to the geometric size of the molecule itself.  
	
	In the present work we have measured the lifetimes of the $\nu=0$, 1, and 2 vibrational states of strontium Rydberg molecules together with the lifetime of the parent $5\textup{s}38\textup{s}~^3\textup{S}_1$ Rydberg atoms.  Strontium is attractive for studies of Rydberg molecules because, at least for the even isotopes, spectral complexities associated with the Rydberg electron spin-orbit and ground-state hyperfine interactions are absent.  Furthermore, there is no p-wave scattering resonance which simplifies the molecular potential curves and eliminates the potential step seen in rubidium (and cesium) permitting more direct comparison between theory and experiment.  In marked contrast to the earlier results for rubidium, we observe that the lifetime of the parent Rydberg atom is essentially equal to that of the $\nu=0$ and $\nu=1$ molecular Rydberg states, although the lifetime of the $\nu=2$ state is somewhat shorter.  We also observe that collisions lead to only a small density dependence in the lifetimes of these states. 
\section{Experimental Method}
The production of long-range Rydberg molecules requires ultracold temperatures to ensure that thermal energies are small compared to the molecular binding energies.  Furthermore, high density samples are required to ensure a sizeable probability for finding two atoms with separations comparable to the size of the parent Rydberg atom.  We satisfy these requirements through use of $^{84}\textup{Sr}$ atoms confined in an optical dipole trap (ODT).  This isotope is chosen because of its collision properties which are favorable for evaporative cooling and for creation of high-phase-space-density samples.
	
	The cooling and trapping of strontium atoms is described in detail elsewhere \cite{2014Stellmer, PhysRevLett.103.200402}. The atoms are held in an ODT formed using crossed $1.064~\mu\textup{m}$ laser beams each with a $e^{-2}$ diameter of $60~\mu\textup{m}$.  The ODT typically contains $10^6$ atoms at a temperature of ${\sim}2~\mu\textup{K}$ resulting in an average density in the trap of ${\sim}1.5{\times}10^{13}~\textup{cm}^{-3}$ which corresponds to an average interparticle spacing of ${\sim}0.4~\mu\textup{m}$.  This spacing is somewhat larger than the size of an $n=38$ atom, ${\sim}2(\textup{n}-\delta)^2$ a.u. (where $\delta$, the quantum defect, is 3.371), i.e., ${\sim}0.13~\mu \textup{m}$.  The Rydberg atoms are created by two-photon excitation via the intermediate $5\textup{s}5\textup{p}~^3\textup{P}_1$ state which requires radiation at 689 and $319~\textup{nm}$.  The $689~\textup{nm}$ laser for the first step is tuned $80~\textup{MHz}$ to the blue of the intermediate state to avoid scattering from the atomic state and from molecular states to the red of the atomic transition \cite{PhysRevA.90.032713}.  The light at $319~\textup{nm}$ is generated by doubling the output of a fiber-based optical parametric oscillator.  The 319 and $689~\textup{nm}$ laser beams cross at right angles and have orthogonal linear polarizations resulting in excitation of $^3\textup{S}_1$ states with $M=\pm1$.  The number of Rydberg atoms/molecules present after excitation is determined using selective field ionization (SFI).  An electric field with a risetime $\tau_\textup{R}=9~\mu\textup{s}$ and peak value $500~\textup{V\,cm}^{-1}$ is applied across the ODT and the number of resulting electrons is determined as a function of time, with a bin width of $100~\textup{ns}$, by directing them to a microchannel plate (MCP).  The population of ground state atoms present in the trap is measured by releasing them from the ODT and, after a fall time of $10~\textup{ms}$, measuring their number by absorption imaging using $461~\textup{nm}$ radiation tuned to the $5\textup{s}^2~^1\textup{S}_0\rightarrow 5\textup{s}5\textup{p}~^1\textup{P}_1$ transition.  The observed trap lifetime is $10~\textup{s}$, and is limited by collisions with background gas.
	
\begin{figure}
\includegraphics{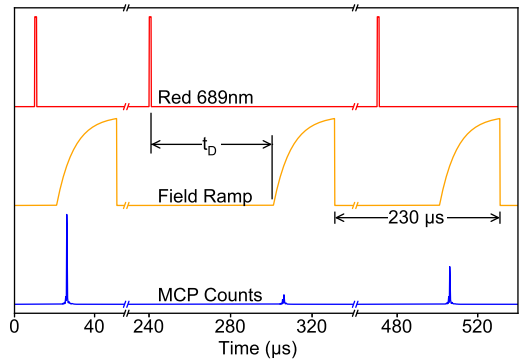}
\caption{\label{fig:timingDiagram} Schematic of the timing sequence used to determine lifetimes.}
\end{figure}

	The lifetimes of the various states are determined by measuring the time rates of decay of their populations using the procedure outlined in Fig.~\ref{fig:timingDiagram}.  A sample of cold atoms is first loaded into the ODT.  Rydberg atoms or molecules are then created using the two-photon scheme discussed previously.  The $319~\textup{nm}$ laser is tuned on resonance with the atomic or molecular line of interest. The output of the $689~\textup{nm}$ laser is chopped to form a periodic train of pulses with a pulse repetition frequency of ${\sim}4~\textup{kHz}$ and pulse durations of $1\textendash{}5~\mu \textup{s}$ chosen to produce $ 1\textendash{}10$ Rydberg atoms or molecules per pulse.  The ODT is turned off during each excitation pulse to eliminate AC Stark shifts.  After some time delay, $t_D$, following the end of the laser pulse the number of surviving Rydberg atoms or molecules is determined by SFI. The $319~\textup{nm}$ laser is turned off during this data acquisition period.
	
	Following initial loading of a cold atom sample into the trap a series of lifetime measurements are undertaken at ${\sim}1.6~\textup{s}$ intervals over a period of ${\sim}10~\textup{s}$. Within each interval, the time delay $t_D$ is cycled repeatedly through a preselected pattern of values ranging from 0 to $60~\mu s$ in $10~\mu s$ intervals and the SFI spectra for each delay time are accumulated. This methodology eliminates systematic errors that might result from slow changes in the ground state atom population in the trap during the course of a lifetime measurement. During each interval, the density only changes by ${\sim} 20\%$ and is treated as a constant for the analysis of our data. In the course of a measurement we sample the density dependence of the atomic/molecular lifetime over a density variation of about an order of magnitude. The results presented here represent the average of data recorded following $150\textendash{}250$ initial trap loadings.
\section{Results and Discussion}
\begin{figure}
\includegraphics{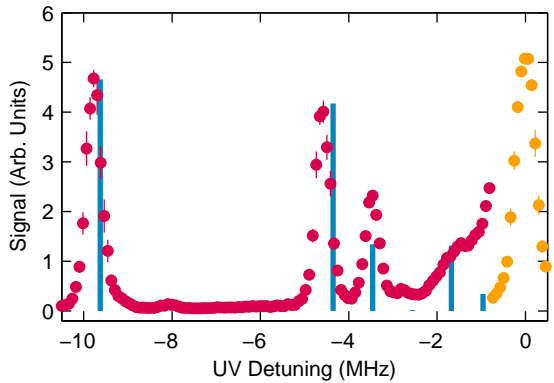}
\caption{\label{fig:UVSpectrum} Measured Rydberg excitation spectrum (the parent $38^3\textup{S}_1$ feature is strongly attenuated). The calculated positions and relative excitation strengths for the  $\nu=0$, 1, and 2 molecular vibrational states are indicated by the vertical bars.  The small additional feature at a binding energy of $8~\textup{MHz}$ is attributed to trimer formation (see text).}
\end{figure}

Figure~\ref{fig:UVSpectrum} shows a representative atom/molecule excitation spectrum.  Three prominent excitation peaks are evident to the red of the parent Rydberg peak.  The widths of these features, $300~\textup{kHz}$, is determined principally by the width of the $319~\textup{nm}$ laser.  The most tightly bound of these features at a detuning of $-9.8~\textup{MHz}$ corresponds to the production of Rydberg (dimer) molecules in the ground $\nu=0$ vibrational state.  The peaks at binding energies of $-4.6$ and $-3.4~\textup{MHz}$ correspond to creation of molecules in the first and second excited vibrational states,  i.e., the $\nu=1$ and $\nu=2$ states, respectively.  

\begin{figure}
\includegraphics[scale = 0.5]{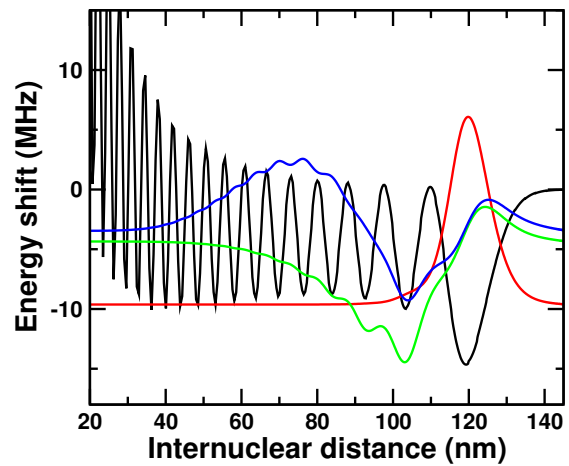}
\caption{\label{fig:wavefunctions} Calculated molecular potential (black) and calculated wave functions for the $\nu=0$ (red), 1 (green), and 2 (blue) molecular vibrational states.}
\end{figure}

	The binding energies can be calculated, as described elsewhere \cite{PhysRevA.92.031403}, with the aid of a two-active-electron (TAE) model and first-order perturbation theory  that utilizes a Fermi pseudopotential and effective s- and p-wave scattering lengths to describe the interaction between the excited Rydberg electron and ground state atom.  The calculated molecular potential within this level of approximation, together with the $\nu=0$, 1, and 2 vibrational wave functions, are shown in Fig.~\ref{fig:wavefunctions}.  The $\nu=0$ wave function is strongly localized at large internuclear separations $R$ in the outermost well of the molecular potential.  The $\nu=1$ wave function extends over several wells and penetrates to somewhat smaller $R$ but is still peaked at relatively large $R$.  The $\nu=2$ wave function is less well localized and extends to relatively small values of $R$.  
	
\begin{figure}
\includegraphics{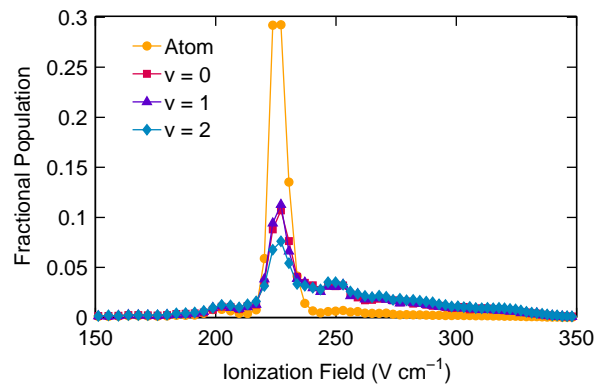}
\caption{\label{fig:DiffStates} SFI spectra measured for the parent $38~^3\textup{S}_1$ Rydberg state and for the $\nu=0$, 1, and 2 molecular states with a delay time $t_D=0$.  The spectra are normalized to the same total area.}
\end{figure}

\begin{figure}
\begin{subfigure}
{\includegraphics{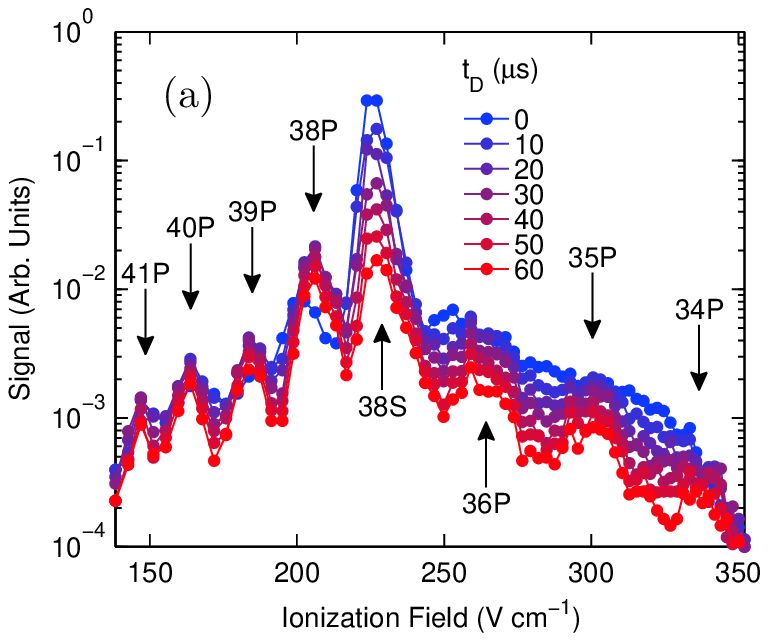}}
\label{fig:SubA}
\end{subfigure}
\begin{subfigure}
{\includegraphics{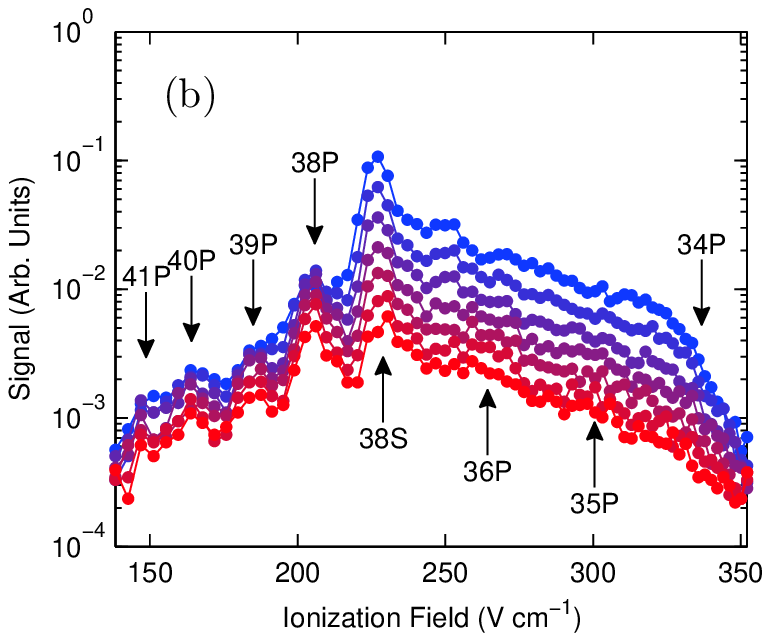}}
\label{fig:SubB}
\end{subfigure}
\caption{\label{fig:VaryDelayTime} Evolution of the SFI spectrum for (a) the parent $38^3\textup{S}_1$ Rydberg state and (b) the $\nu=0$ molecular vibrational state as a function of delay time.  The values of the various delay times used are as indicated.  The arrows indicate the $n$ values of $^3\textup{P}$ states populated by BBR-induced transitions.}
\end{figure}

\begin{figure}
\includegraphics{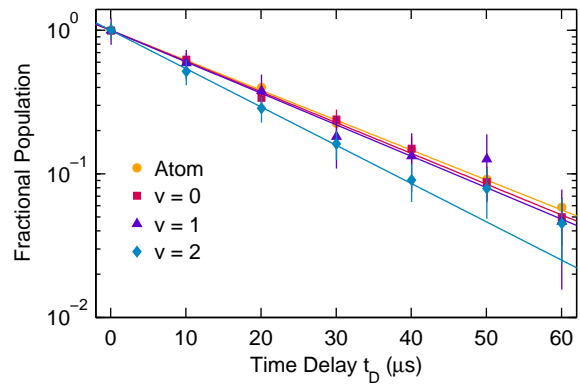}
\caption{\label{fig:StateEvolution} Evolution of the population of $38^3\textup{S}_1$ Rydberg atoms and of the molecular $\nu=0$, 1, and 2, vibrational states as a function of time delay $t_D$.   The data sets are each normalized to one at $t_D=0$.}
\end{figure}

\begin{figure}
\includegraphics{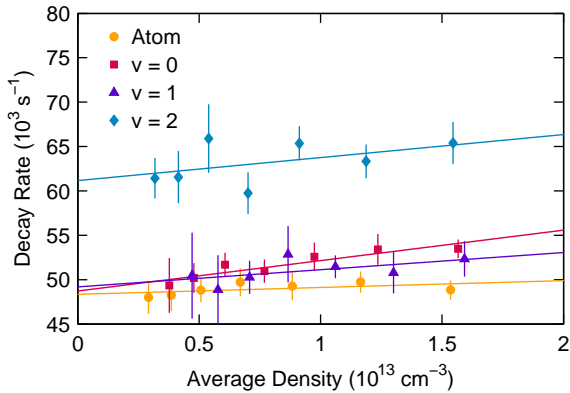}
\caption{\label{fig:DecayvsDensity} Measured decay rates $\gamma_{\textup{TOT}}$ as a function of average trap density.  Straight line fits to the data are included.}
\end{figure}

	The calculated molecular binding energies shown in Fig.~\ref{fig:UVSpectrum} are in good agreement with the measured values.  The calculated relative excitation strengths are also shown and are in reasonable accord with the data.  Additional more-weakly-bound dimer states are predicted and visible in the spectrum but their small signal strengths precluded measurement of their lifetimes. The small feature seen at a binding energy of ${\sim}8.0~\textup{MHz}$ is most likely due to excitation of a trimer state with two bound atoms, one in the $\nu=1$ state, the other the $\nu=2$ state.
\break \indent Typical SFI spectra recorded for the $5\textup{s}38\textup{s}~^3\textup{S}_1$ atomic Rydberg state and the $\nu=0$, 1, and 2 molecular states are shown in Fig.~\ref{fig:DiffStates}.  A delay time $t_D=0$ was used to minimize the effects of blackbody\textendash{}radiation(BBR)-induced transitions to neighboring levels.  The data sets are normalized to the total number of atoms/molecules to emphasize the differences in their shapes.  The atomic spectrum peaks at a field of $230~\textup{V\,cm}^{-1}$ consistent with the adiabatic ionization of $38^3\textup{S}_1$ states for which the threshold field is given approximately by $1/16(n-\delta)^4$.  Interestingly, the SFI spectra for the molecular states are broader than that of the parent Rydberg atom.  While the origin of this difference is not the primary focus of this paper, attempts to explain the results by expanding the molecular electronic wavefunction in terms of atomic states with different values of $n, l$, and $m$, and which ionize at different field strengths, were unsuccessful because the contributions from such states proved to be negligibly small \cite{PhysRevLett.115.023001}.  The data, therefore, suggest that electron scattering from the neutral atom during the ionizing field ramp influences the ionization process, a hypothesis that might be tested in the future by pushing to higher trap densities and by measuring SFI spectra for trimer and tetramer Rydberg molecules.
\break	\indent The evolution of the SFI spectra with increasing delay time is shown, on a logarithmic scale, in Fig.~\ref{fig:VaryDelayTime}.  For the parent $38^3\textup{S}_1$ atoms the size of the main central peak decreases rapidly with increasing delay time $t_D$.  At the same time several pronounced features appear on both the high and low field sides of the parent feature.  These peaks are associated principally with population of neighboring $^3\textup{P}$ states through BBR-induced transitions.  The corresponding values of $n$ are indicated in the figure.  Since P states are being continuously generated from the parent state and since TAE calculations indicate that their lifetimes are approximately three times longer than that of the parent state, their populations decrease less rapidly than that of the parent state.  As shown in Fig.~\ref{fig:VaryDelayTime}, similar behavior is seen for the $\nu=0$ molecular state although the high-field BBR-induced features are more difficult to discern due to the sizeable high-field shoulder present in the SFI spectra.
	\break \indent Lifetimes were determined by observing the time dependence of the population in the small central region of the dominant peak in each spectrum which is proportional to the number of parent Rydberg atoms/molecules remaining (Fig.~\ref{fig:StateEvolution}). The voltage range of the integration was chosen to minimize the contribution to the integrated signals from the underlying P states while maintaining a reasonable signal level.  Tests revealed, however, that expanding the integration range to encompass each entire peak resulted in less than a 6\% reduction in the measured decay rates. Figure~\ref{fig:StateEvolution} shows the time evolution of the number of surviving parent $38^3\textup{S}_1$ Rydberg atoms and molecular $\nu=0$, 1, and 2, vibrational states for an average cold atom density of ${\sim}0.4{\times}10^{13}~\text{cm}^{-3}$. Each of these populations decays essentially exponentially and the corresponding decay rates $\gamma_{\textup{TOT}}$ are plotted in Fig.~\ref{fig:DecayvsDensity} together with those measured at other trap densities.  
	
Loss of parent Rydberg atoms/molecules from the main peak can result from radiative decay, from BBR-induced transitions to neighboring states, and from collisions.  The total decay rate is therefore given by		
\begin{equation}
\gamma_{\textup{TOT}}= \gamma_{\textup{R}}+ \gamma_{\textup{BBR}}+ \gamma_{\textup{COLL}}
\end{equation}
where $\gamma_\textup{R}$ is the radiative decay rate, $\gamma_{\textup{BBR}}$ the BBR-induced decay rate, and $\gamma_{\textup{COLL}}$ the rate for collisional loss through, for example, $n$- and $l$-changing collisions.  The collisional loss rate $\gamma_{\textup{COLL}}$ can be written as
\begin{equation}
\gamma_{\textup{COLL}}=\rho k_{\textup{COLL}}=\rho \sigma_{\textup{COLL}} \left \langle v \right \rangle
\end{equation}
where $\rho$ is the trap density, $k_{\textup{COLL}}$ ($\sigma_{\textup{COLL}}$) is the rate constant (cross section) for collisional loss, and $\left \langle v \right \rangle$  is the average relative collision velocity taken to be the mean atom-atom collision velocity, ${\sim}3.2{\times}10^{-2}~\textup{ms}^{-1}$.

	The decay rate for the $38^3\textup{S}_1$ Rydberg state displays little, if any, dependence on trap density demonstrating that  the collisional loss cross section $\sigma_{\textup{COLL}}= \left \langle v \right \rangle^{-1}d\gamma_{\textup{TOT}}/d\rho$ is negligible. The results in Fig.~\ref{fig:DecayvsDensity} yield a decay rate $\gamma_{\textup{TOT}}=4.8(2){\times}10^4~\textup{s}^{-1}$ corresponding to an effective atomic lifetime $\gamma_{\textup{TOT}}^{-1}=21(1)~\mu\textup{s}$.
\break \indent Near exponential decreases are also observed in the number of surviving parent  $\nu=0$, 1 and 2 molecular states.  The corresponding decay rates $\gamma_{\textup{TOT}}$ are included in Fig.~\ref{fig:DecayvsDensity} and exhibit a weak dependence on trap density.  While the experimental uncertainties make it difficult to obtain an accurate measure of this dependence, the data suggest that the density dependences are similar for all the vibrational levels and point to an “averaged” collisional loss cross section $\sigma_{\textup{COLL}}\sim8{\times}10^{-11}~\textup{cm}^2$.   

	Collisional loss is not unexpected for molecules given that an additional destruction channel is available, namely collisions involving the weakly-bound neutral atom.  This cross section for collisional loss, however, is smaller than the geometric sizes of the molecules inferred from Fig.~\ref{fig:wavefunctions}, ${\sim}3\textendash{}5{\times}10^{-10}~\textup{cm}^2$.  This contrasts with the behavior seen with rubidium $35\textup{s}~^2\textup{S}_{1/2}$ molecules where the cross section for collisional destruction was comparable to the geometrical size of the molecule. A clear understanding of the mechanisms leading to the destruction of Rydberg molecules is lacking but our measurements suggest that, for rubidium, the strong p-wave electron scattering resonance plays an important role.
\break \indent	Extrapolation of the data in Fig.~\ref{fig:DecayvsDensity} to zero trap density yields $\gamma_{\textup{TOT}}=4.9(4)$, 4.9(5), and $6.1(4){\times}10^4~\textup{s}^{-1}$ for the $\nu=0$, 1, and 2 molecular states, respectively, corresponding to effective lifetimes of 20(2), 20(2), and $16(2)~\mu s$.   While the lifetimes of the parent Rydberg atoms and the molecular $\nu=0$ and 1 states are very similar, that of the $\nu=2$ molecular state is ${\sim}20\%$ shorter.  This behavior contrasts markedly with that seen for rubidium, and it can be explained by noting that, as shown in Fig.~\ref{fig:wavefunctions}, for strontium the lack of a p-wave resonance leads to a sizeable repulsive barrier in the molecular potential that limits the close approach of the constituent atoms.  The $\nu=2$  molecular state does, however, penetrate closer to the Rydberg core than do the lower-lying states which might open up additional short-range loss channels.  At small internuclear separations simple perturbation theory based on a Fermi pseudopotential is no longer appropriate and calculations become more difficult.	
\break \indent	Since BBR-induced transitions populate neighboring Rydberg states, the decay of the total population in all Rydberg levels is governed simply by radiative decay and collisions. Given the relatively long lifetime of the $^3\textup{P}$ states populated by BBR, radiative decay of the $38 ^3\textup{S}_1$ state will, at least at early times, be the dominant radiative loss process and the overall decay rate can be written approximately as 
\begin{equation}
\gamma_{\textup{TOT}}'\approx\gamma_{\textup{R}}+\gamma_{\textup{COLL}}.
\end{equation}
Measurements of the decay of the total Rydberg populations undertaken by integrating over the full electric field range shown in Fig.~\ref{fig:VaryDelayTime} again revealed near exponential decay.  For the $38^3\textup{S}_1$ atoms $\gamma_{\textup{TOT}}'= 3.1{\times}10^4~\textup{s}^{-1}$ yielding the BBR-induced decay rate,  $\gamma_{\textup{BBR}}\approx\gamma_{\textup{TOT}}-\gamma_{\textup{TOT}}'$, ${\sim} 1.7{\times}10^4~\textup{s}^{-1}$, a value in good agreement with that calculated using the TAE model, $\gamma_{\textup{BBR}}=1.4{\times}10^4~\textup{s}^{-1}$, and one that is consistent with those reported previously in earlier studies of interactions between BBR and Rydberg atoms \cite{PhysRevA.75.052720, 0953-4075-43-12-125002, 0953-4075-44-19-195010}.  The effective lifetime of $20~\mu\textup{s}$ calculated using the TAE model is also in good agreement with the measured value of $21(1)~\mu\textup{s}$, although both values are substantially larger than the value reported previously for the $35^3\text{S}_1$ state of $7.5(44)~\mu\textup{s}$ \cite{kunze1993lifetime}.  The difference in the decay rates  $\gamma_{\textup{TOT}}$  and  $\gamma_{\textup{TOT}}'$ for the molecular states were similar to that for the parent atoms, namely 1.7, 1.7, and $2.2 {\times}10^4~\textup{s}^{-1}$  for the $\nu=0$, 1, 2 states, respectively.
\break \indent	In summary, the present measurements show that, as compared to rubidium, the lifetimes of the low-lying vibrational states in ultralong-range strontium Rydberg molecules are much closer to those of the parent Rydberg atoms and display a much weaker dependence on the vibrational quantum number $\nu$.  This demonstrates the importance of the p-wave scattering resonance (or lack thereof) in governing the behavior of Rydberg atoms and molecules in dense gases. Additionally, the lifetimes of the strontium molecular states are much less sensitive to collisions with ground state atoms. This work suggests that the strontium system with its relative simplicity offers advantages for testing theories of molecular formation, structure, and collisional loss.

\begin{acknowledgments}
Research supported by the AFOSR under Grant No. FA9550-14-1-0007, the NSF under Grants No. 1301773 and No. 1205946, the Robert A, Welch Foundation under Grants No. C-0734 and No. C-1844, the FWF(Austria) under Grant No. P23359-N16, and FWF-SFB049 NextLite.  The Vienna scientific cluster was used for the calculations.  HRS was supported by a grant to ITAMP from the NSF.
\end{acknowledgments}

%

\end{document}